\documentclass{aa}

\usepackage{graphicx,natbib}

\newcommand{\micron}{\ensuremath{\mu\mbox{m }}}
\bibpunct{(}{)}{;}{a}{}{,}

\begin{document}
\thesaurus{03(03.09.2)}
\title{Integrated optics for astronomical interferometry}
\subtitle{II. First laboratory white-light interferograms}
\author{J.P. Berger\inst{1}
  \and K. Rousselet-Perraut\inst{1}
  \and P. Kern\inst{1}
  \and F. Malbet\inst{1}
  \and I. Schanen-Duport\inst{2}
  \and F. Reynaud \inst{3}
  \and P. Haguenauer \inst{1, 4}
  \and P. Benech\inst{2}}
\institute{ 
  Laboratoire d'Astrophysique UMR CNRS/UJF 5571, Observatoire de Grenoble,
  BP 53, F-38041 Grenoble cedex 9, France
  \and
  Laboratoire d'\'Electromagn\'etisme Microondes et Opto\'electronique UMR
  CNRS/INPG/UJF 5530, BP 257, F-38016 Grenoble cedex 1, France
  \and
  Institut de Recherche en Communications Optiques et Microondes, UMR
  CNRS/Univ.\ Limoges 6615, 123 avenue Albert Thomas, F-87060 LIMOGES
  cedex, France
  \and 
  CSO Mesures 70, Avenue des Martyrs, F-38000 Grenoble, France}
\offprints{J.P. Berger}
\mail{berger,malbet@obs.ujf-grenoble.fr}
\date{Received February 25; Accepted May, 1999\hfill 
  \textit{Astron.\ Astrophys.\ Suppl.\ Ser.\ } in press (1999)}
\authorrunning{J.-P. Berger et al.}
\titlerunning{\small Integrated optics for interferometry. II.
 Lab white-light interferograms} 
\maketitle

\begin{abstract}
  We report first white-light interferograms obtained with an
  integrated optics beam combiner on a glass plate. These results
  demonstrate the feasability of single-mode interferometric beam
  combination with integrated optics technology presented and
  discussed in paper I \citep{Mal99}. The demonstration is
  achieved in laboratory with off-the-shelves components coming from
  micro-sensor applications, not optimized for astronomical use. These
  two-telescope beam combiners made by ion exchange technique on glass
  substrate provide laboratory white-light interferograms
  simultaneously with photometric calibration. A dedicated
  interferometric workbench using optical fibers is set up to
  characterize these devices. Despite the rather low match of the
  component parameters to astronomical constraints, we obtain
  stable contrasts higher than 93$\%$ with a 1.54-\micron laser source 
  and up to 78$\%$ with a white-light source in the astronomical H band. 
  Global throughput of 27$\%$ for a potassium ion exchange beam combiner 
  and of 43$\%$ for a silver one are reached. This work validates 
  our approach for combining several stellar beams of a long baseline 
  interferometer with integrated optics components.
\end{abstract}


\section{Introduction}
\label{sect:intro}  

Since \citet{Fro81} proposed guided optics for astronomical interferometry,
important progress has been made.  In particular, the FLUOR instrument
which combines two interferometric beams with single-mode fiber couplers
\citep{Cou94} has led to astrophysical results with unprecedent precision.
This experiment demonstrated the great interest of spatial filtering
combined with photometric calibration to improve the visibility accuracy.
More recently, \citet{Ker96} suggested to combine interferometric beams
with integrated optics components since this technology allows to
manufacture single-mode waveguides in a planar substrate. In paper I
\citep{Mal99}, we presented and discussed thoroughly the advantages and
limitations of integrated optics for astronomical interferometry. To
validate the latter analysis, we have performed several laboratory
experiments with existing components not optimized for interferometry but
allowing to get first clues on this technology. The first experimental
results are reported in this Letter.

\section{Experimental set-up}
\label{sect:setup}

\begin{figure*}[t]
  \begin{center}
    \leavevmode
    \includegraphics[angle=-90,width=0.95\textwidth]{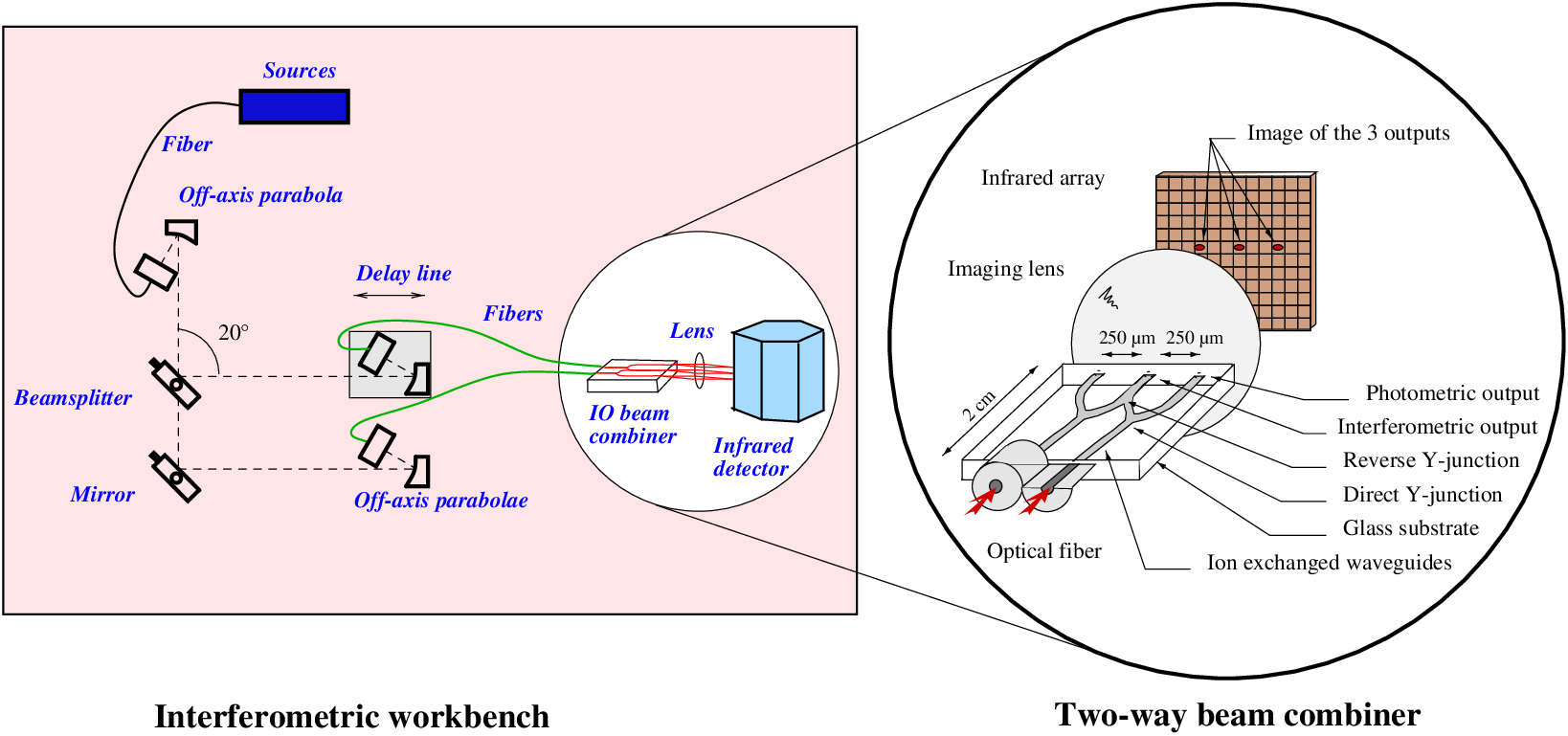}
    \caption{Left: laboratory interferometric workbench for testing integrated
      optics beam combiners based on a Mach-Zehnder interferometer concept.
      The collimated beam provided by the sources (laser, laser diode or
      white-light source) is splitted in two beams which are focused onto
      low-birefringence single-mode fibers by off-axis parabolae.  Fibers
      are directly connected to the integrated optics beam combiner.  The
      optical path difference between the beams is modulated by a
      translating stage to produce a delay. The three outputs of the
      beam combiner are imaged on a cooled infrared HgCdTe array. The angle
      between the incident and reflected beams on the beamsplitter is close
      to $20^\circ$.
      Right: integrated optics two-way beam combiner. The light 
      from the telescopes is coupled into the integrated optics beam
      combiner thanks to optical fibers. Two direct Y-junctions provide
      photometric calibration signals for each beam. A reverse Y-junction
      combines the two input beams. Each output is imaged by a lens onto an
      infrared array.}
    \label{fig:experiment}
  \end{center}
\end{figure*}

We carried out laboratory tests with off-the-shelves integrated optics
components designed for micro-sensor applications. The waveguides are made
by ion exchange (here potassium or silver) on a standard glass substrate
thanks to photo\-lithography techniques \citep{Sch96}. The exchanged
area is analogous to the core of an optical fiber and the glass substrate
to the fiber cladding. Our $5 \mbox{ mm} \times 40 \mbox{ mm}$ component is
schematically depicted in the right part of Fig.\ \ref{fig:experiment}.  We
use it as a two-way beam combiner with two photometric calibration signals.
The component operates in the H atmospheric band (1.43\micron -
1.77\micron) and its waveguides are single-mode in that domain.  From an
optical point of view, the reverse Y-junction acts as one of the two
outputs of a classical beam splitter. The second part of the
interferometric signal with a $\pi$ phase shift is radiated at large scale
in the substrate. Light is carried to the component thanks to standard
non-birefringent silica fibers.

We have set up a laboratory Mach-Zehnder interferometer to test the
interferometric capabilities of our components (see the left part of Fig.\ 
\ref{fig:experiment}).  The available sources are: a 1.54-\micron He-Ne
laser, a 1.55-\micron laser diode and an halogen white-light source. The
latter is used with an astronomical H filter.

We scan the interferograms by modulating the optical path difference (OPD)
with four points per fringe. The delay line speed is restricted by the
integration time ($\sim$1 ms for laser sources and $\sim$10 ms for the
white-light source to get a sufficient signal-to-noise ratio) and the frame
rate (50 ms of read-out time for the full frame). The OPD scan and the data
acquisition are not synchronized, but for each image the translating stage
provides a position with an accuracy of 0.3\micron.  The simultaneous
recording of the photometric and interferometric signals allows to unbias
the fringe contrast from the photometric fluctuations as suggested by
\citet{Con84} and validated by \citet{Cou94}.

\begin{figure*}[t]
  \begin{center}
    \leavevmode
    \hfill
    \includegraphics[width=0.3\textwidth]{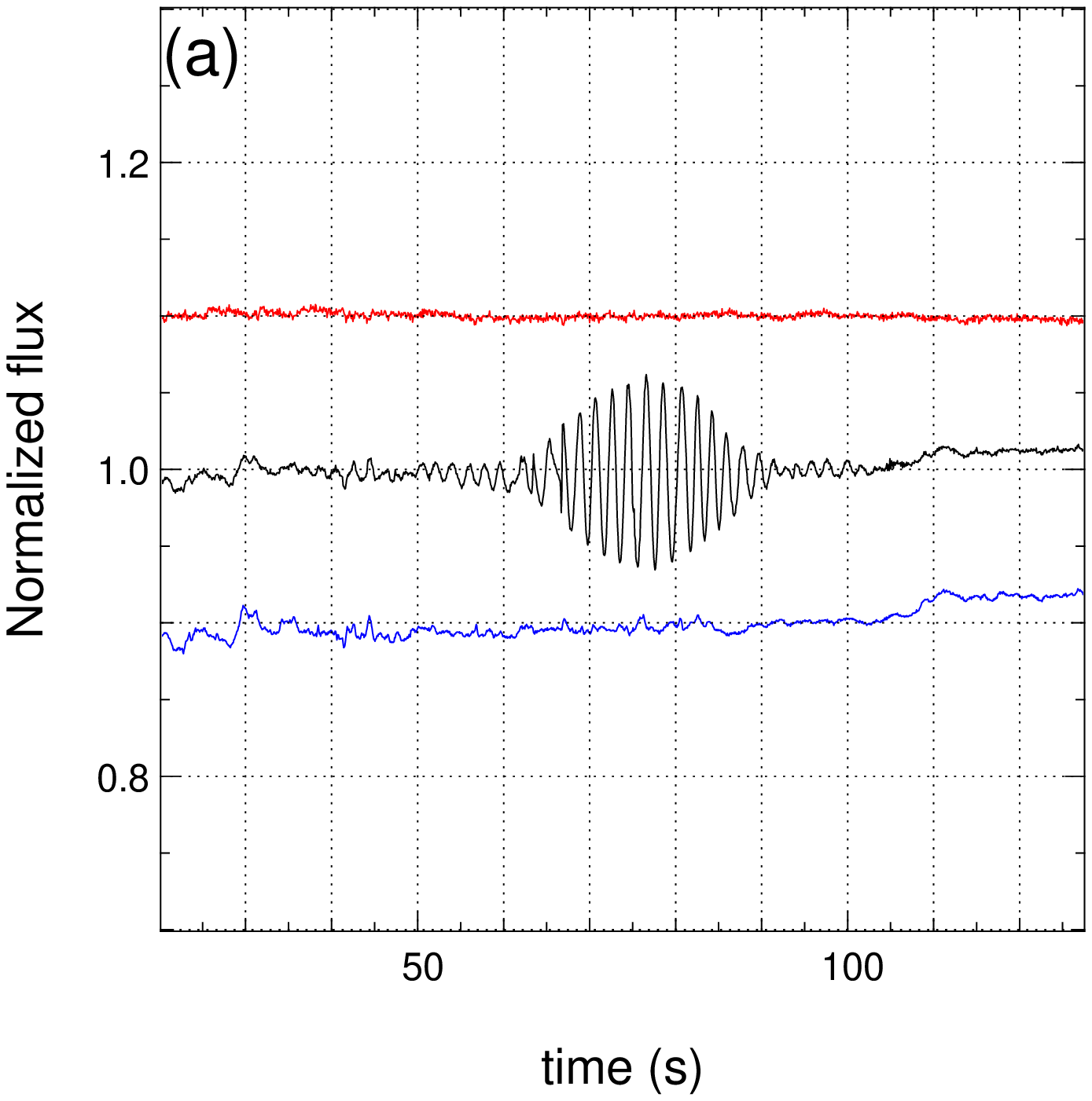}
    \hfill
    \includegraphics[width=0.3\textwidth]{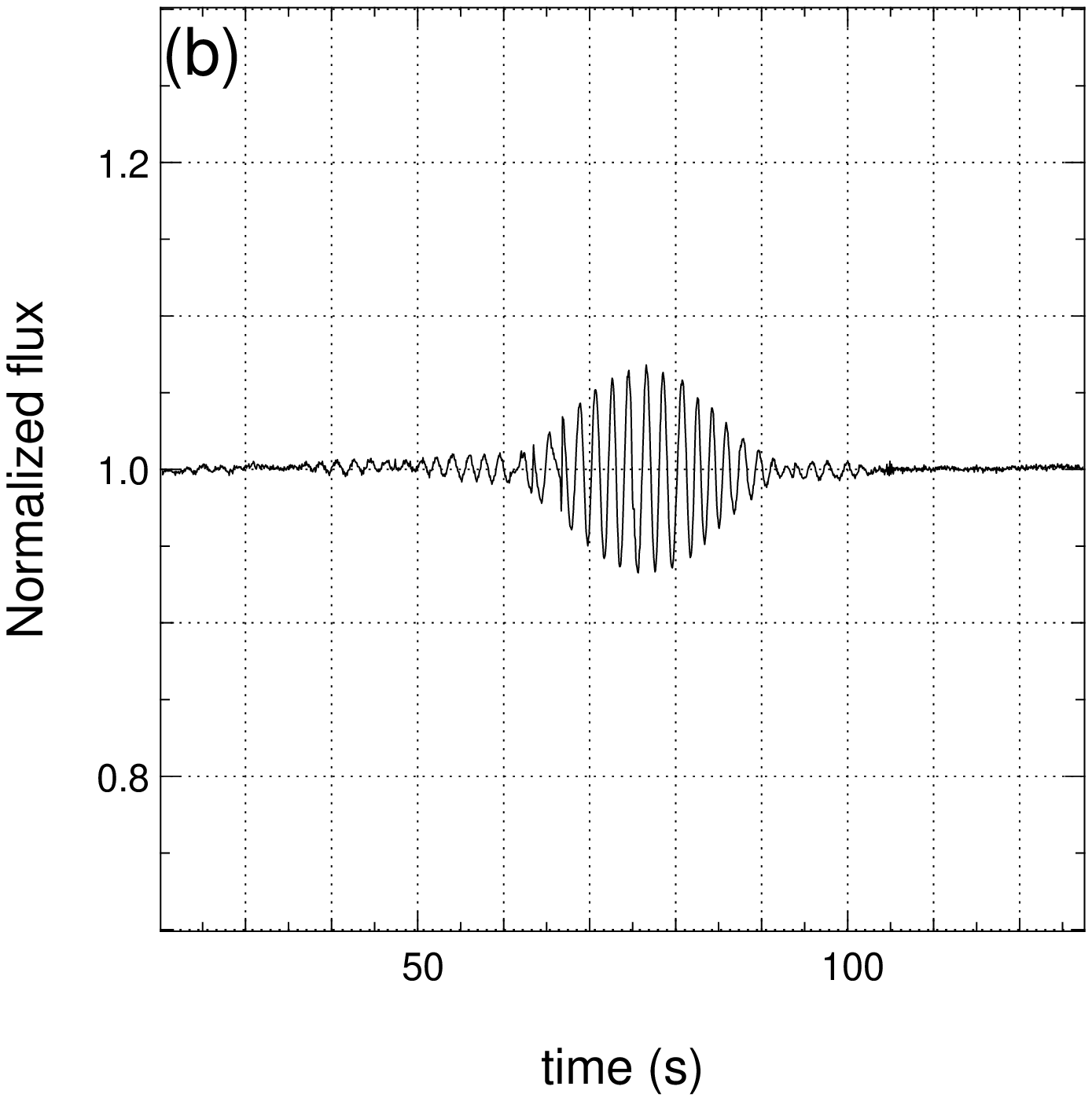}
    \hfill
    \includegraphics[width=0.3\textwidth]{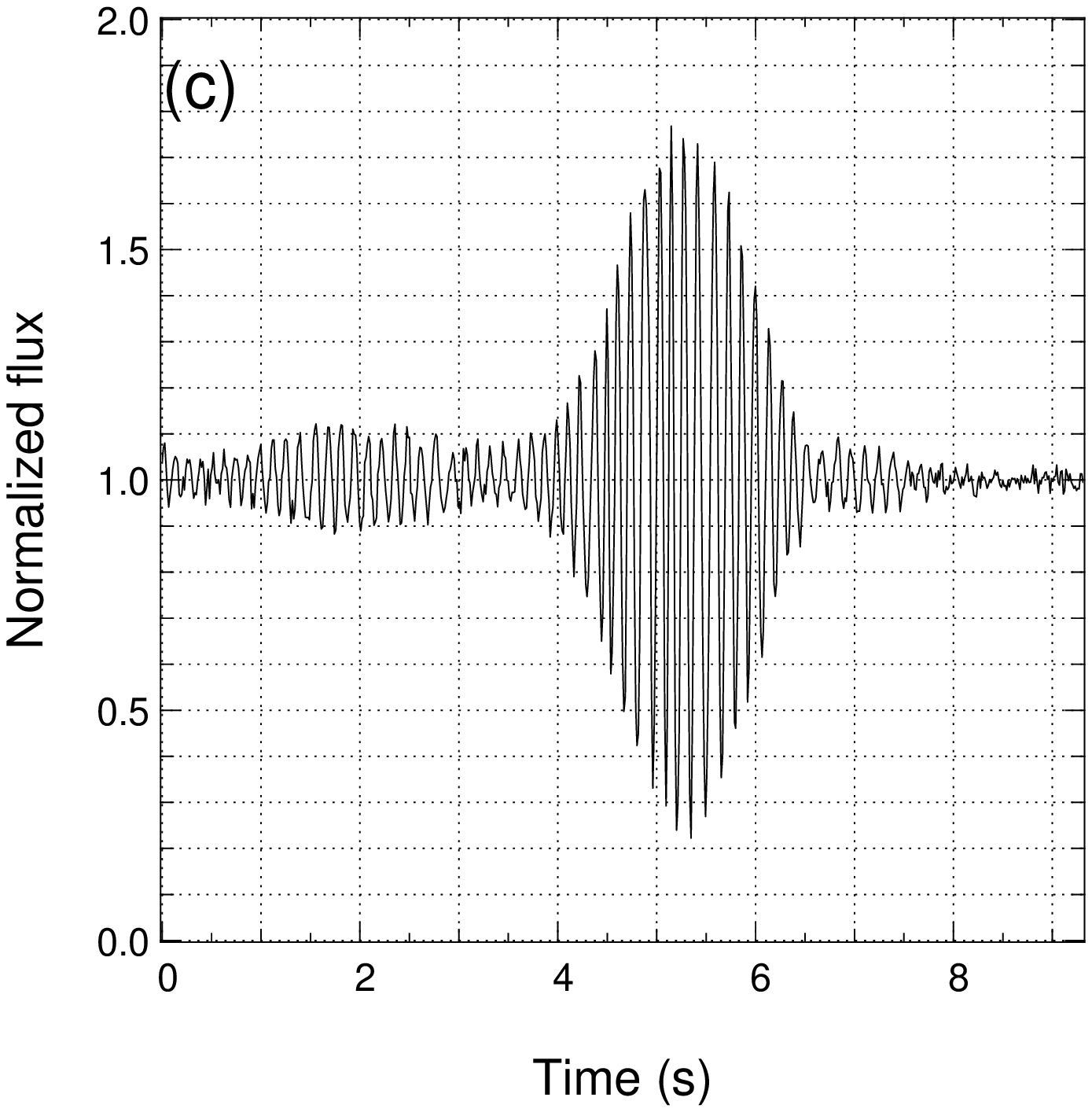}
    \hfill~\\
    \caption{White-light interferograms obtained in laboratory with 
      integrated optics components: potassium ion exchange component
      connected with low-birefringent fibers (a, b) and silver ion exchange
      component connected with high-birefringent fibers (c). (a) Raw
      interferogram in the H atmospheric band with photometric
      calibration signals (upper and lower curves with vertical shifts of
      $+0.1$ and $-0.1$) in normalized units. (b) Interferogram corrected from
      the photometry (b) in same units. (c) Photometry-corrected interferogram
      obtained with input beams polarized along the neutral axis of
      high-birefringent fibers.}
    \label{fig:interf}
  \end{center}
\end{figure*}

A typical white light interferogram $I_{0}$ is plotted in Fig.\ 
\ref{fig:interf}a together with the simultaneous photometric signals
$P_{1}$ and $P_{2}$. To correct the raw interferogram from the photometric
fluctuations, we substract a linear combination of $P_{1}$ and $P_{2}$ from
$I_{0}$. The expression of the corrected interferogram is then
\begin{equation}
I_{c}=\frac{I_{0}-\alpha P_{1}-\beta P_{2}}{2\sqrt{\alpha P_{1}\,\beta P_{2}}}
\end{equation}
with $\alpha$ and $\beta$ coefficients determined by occulting
alternatively each input beam.  The resulting corrected interferogram is
displayed in Fig.\ \ref{fig:interf}b.

\section{Results and discussion}
\label{sect:res}

As far as an astronomer is concerned, the physical quantities of interest
when dealing with an interferometric instrument are the instrumental
contrast, the optical stability, and the total optical throughput.

\subsection{Laser-light constrasts}

A 93\% contrast is obtained with the He-Ne laser. The main source of
contrast variations with time comes from temperature gradient and
mechanical constraints on the input fibers. When special care is taken to
avoid fiber bending and twists, the laser contrast variation is lower than
7$\%$ rms over a week. Using high-birefringent fibers which are far less
sensitive to mechanical stresses will improve the contrast stability.

\subsection{White-light contrasts}

With an halogen white-light source, the contrast obtained is of the order
of 7$\%$ with a potassium ion beam combiner connected with low-birefringent
fibers (Fig. \ref{fig:interf}b) and 78$\%$ with a silver ion beam combiner
connected with high-birefringent fibers (Fig. \ref{fig:interf}c). Two main
sources of interferometric contrast drops between the two components have
been identified: chromatic dispersion and polarization mismatch.

The consequence of residual differential dispersion between the two arms is
to spread out the fringe envelope and decrease the contrast.  Since the
delay line translation is not perfectly linear, the Fourier relation
between space and time is affected and an accurate estimate from the
dispersion is difficult. Only the number of fringes and the shape of the
interferogram gives an idea of the existing differential dispersion. The
theoretical number of fringes is given by the formula
$2\frac{\lambda}{\Delta \lambda}\sim 10$ and the interferogram contains
about $13$ fringes. Such a spread is not sufficient to explain the contrast
drop between the laser- and white-light contrasts. More detailed studies of
residual chromatic dispersion are in progress.

In the present case the contrast decay is mainly explai\-ned by differential
birefringence. Low-birefringent fibers are known to be highly sensitive to
mechanical constraints and temperature changes, leading to unpredictable
birefringence. Coupling between polarisation modes can occur leading to a
contrast loss which can be significant for unpolarized incident light (case
of the white-light source). This is confirmed by the preliminary results
obtained with high-birefringent fibers and the incident light polarized
along the neutral axes: the contrast reaches 78$\%$ (Fig.
\ref{fig:interf}c). The apparent asymmetry of the interferogram could be
due to residual differential polarization and/or dispersion. Full
characterizations are in progress and will be reported in Paper III
\citep{Hag99}.

\subsection{Total throughput}

\begin{figure}[t]
  \begin{center} 
    \leavevmode
    \includegraphics[angle=-90,width=0.9\columnwidth]{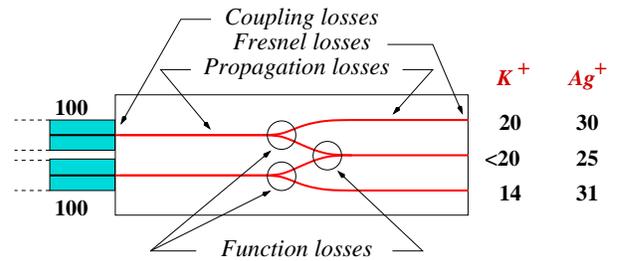}
    \caption{Schematic view of the beam combiner with
      successives optical losses (see text for details). Number of output
      photons are given for 100 incoherent photons injected in each channel
      for potassium- and silver-exchanged waveguides.}
    \label{fig:loss} 
  \end{center}
\end{figure}

Fig. ~\ref{fig:loss} summarizes the photon losses in the two components.
We express the losses in terms of remaining photons when 100 {\em
  incoherent} photons are injected at each waveguide input. For the
component made from potassium ion exchange, we obtain 20 and 14 photons on
each photometric channel and less than 20 photons in the interferometric
channel leading to a total of 54 photons for 200 photons injected, hence a
total throughput of 27\%.  For silver ion exchange, respectively 30, 31 and
25 photons have been measured leading to a throughput of 43\%.  The main
difference between the two results comes from the coupling efficiency
between the fiber and the waveguides and the propagation losses.

\begin{table}[t]
  \caption{Estimation of optical losses at different levels of 
    fiber-connected beam combiners with respectively potassium- and 
    silver-exchanged waveguides. The number of detected and estimated output 
    photons are given for our 4-cm components. Last column gives an idea of 
    what performances can be achieved in the future.} 
  \label{tab:loss}
  \medskip
  \begin{tabular}{ll@{~~~}l@{~~~}l}
    \hline
    Component                      &K$^+$     &Ag$^+$    &\emph{Future}\\
    \hline                         
    \hline                         
    Fiber/waveguide coupling       &40 \%     &20\%      &\emph{4.5\%}\\
    Propagation$^a$                &24\%      &9\%       &\emph{9\%}\\
    Function                       &10\%      &10\%      &\emph{5\%}\\
    Fresnel reflexion              &4\%       &4\%       &-\\
    Beam combination$^b$           &50\%      &50\%      &\emph{-}\\
    \hline                         
    \hline                         
    \emph{Number of input photons} &200       &200       &\emph{-}\\
    Detected photons               &54        &86        &\emph{-}\\
    \hline
    \hline
    Experimental throughput        &27\%      &43\%      &\emph{-}\\
    Theoretical throughput         &28\%      &46\%      &\emph{74\%}\\
    \hline
    \multicolumn{4}{p{0.8\columnwidth}}{\footnotesize $^a$ K$^+$: 0.3
    dB/cm, Ag$^+$: 0.1 dB/cm for a total length of 4cm.} \\
    \multicolumn{4}{p{0.8\columnwidth}}{\footnotesize $^b$ 50\% of the 
    flux in a reverse Y-junction is radiated out in the substrate (see paper I).} \\ 
\end{tabular}
\end{table}

Table \ref{tab:loss} summarizes estimation of losses coming from different
origins. The propagation losses and the coupling losses have been measured
with a straight waveguide manufactured in the same conditions.  The Fresnel
losses have been theoretically estimated to 4$\%$. Any function causes
additional losses which cannot be evaluated separately but have been
estimated to 10$\%$. One should notice that the reverse Y-junction acts as
only one of the two outputs of an optical beamsplitter (see paper I).
Therefore 50$\%$ of the light is radiated outside the waveguide. The first
two columns of Table \ref{tab:loss} show that our measurements are
consistent with the theoretical performances computed from the different
optical losses reported in the Table.

Last column of Table \ref{tab:loss} gives an order of magnitude of expected
improvement in the future. The main progress concerns the beam combination
function. We should be able to retrieve the second half of the combined
photons thanks to new combination schemes like X-couplers, multiaxial beam
combiners or multimode interferometric (MMI) multiplexers (see paper I) at
the cost of a slight chromaticity of the function. Some components
including these new functions are being manufactured and will be soon
tested. The ultimate optical throughput would be around 70-80\%, twice more
than our current results.

\section{Conclusion and future prospects}

We have obtained first high-contrast white-light interferograms with an
off-the-shelves integrated optics component used as a two aperture beam
combiner. The high and stable contrasts as well as the high optical
throughput validate our approach for combining stellar beams by means of
integrated optics presented in paper I.

This preliminary analysis requires further characterizations and
improvements. The importance of dispersion, birefringen\-ce and other
phenomena in the fibers and in the components have to be fully understood.
For this purpose, two-way beam combiners optimized for astronomy are under
characterization (spectroscopic and polarimetric measurements for instance)
in order to carefully control their optical properties. A complete
description of optical and interferometric properties of integrated optics
component will be presented in a forthcoming paper \citep{Hag99}.  The
optical fibers should maintain polarization to avoid specific contrast
losses and have optimized lengths to avoid chromatic dispersion. This
experimental precaution is decisive to achieve image reconstruction
\citep{Del98}.

Our research program is based on the study of new integrated optics
technologies for long baseline interferometry in the infrared, and,
the design of beam combiners for multiple apertures\footnote{Actually
3-way and 4-way beam combiners have already been manufactured} (see
paper I). Some specific beam combiners will then be eventually used in
a scientific instrumental prototype on astronomical interferometers.
Preliminary tests on the GI2T/Regain interferometer \citep{Mou98} will
be carried out with the Integrated Optics Near-infrared
Interferometric Camera (IONIC) prototype \citep{Ber98}.

\section{Acknowledgments}

We would like to warmly thank E.\ Le Coarer for his precious support in
instrument control. We thank the referee, Dr.\ Shaklan, for a careful
reading of our paper and for suggestions which helped to improve its
content. The work was partially funded by PNHRA / INSU, CNRS / Ultima\-tech
and DGA / DRET (Contract 971091). The integrated optics components have
been manufactured and fiber-connected by the GeeO company.

\end{document}